# Mechanism of carrier localization in doped perovskite nanocrystals for bright emission

Sascha Feldmann,[1] Mahesh Gangishetty,[2] Ivona Bravić,[1] Timo Neumann,[1,3] Bo Peng,[1] Thomas Winkler,[1] Richard H. Friend,[1] Bartomeu Monserrat,[1,4] Daniel N. Congreve,[2] and Felix Deschler[1,3*]

[1] *Cavendish Laboratory, University of Cambridge, Cambridge, CB30HE, UK*

[2] *Rowland Institute, Harvard University, Cambridge, Massachusetts, 02142, USA*

[3] *Walter Schottky Institute, Technical University of Munich, Garching, 85748, Germany*

[4] *Department of Materials Science and Metallurgy, University of Cambridge, Cambridge, CB30FS, UK*

*e-mail: felix.deschler@wsi.tum.de

**Abstract**

Nanocrystals based on metal-halide perovskites offer a promising material platform for highly efficient lighting. Using transient optical spectroscopy, we study excitation recombination dynamics in manganese-doped $CsPb(Cl,Br)_3$ perovskite nanocrystals. We find an increase in the intrinsic excitonic radiative recombination rate upon doping, which is typically a challenging material property to tailor. Supported by *ab initio* calculations, we can attribute the enhanced emission rates to increased exciton localization through lattice periodicity breaking from Mn dopants, which increases exciton effective masses and overlap of electron and hole wavefunctions and thus the oscillator strength. Our report of a fundamental strategy for improving luminescence efficiencies in perovskite nanocrystals will be valuable for maximizing efficiencies in light-emitting applications.

**Introduction**

Metal-halide perovskite nanocrystals (NCs) have been the subject of intense investigation due to their high brightness, spectral tunability and excellent color gamut, making them ideal candidates for low-cost and highly efficient light-emitting diodes (LEDs) [1–3]. Recently, doping of these NCs with manganese ions has resulted in increased efficiencies in perovskite-based LEDs [4].

Doping of traditional II-VI, II-V, and group-IV nanocrystal semiconductors was shown to successfully modify electronic, optical and magnetic properties [5–7]. Here, intentionally introduced impurities with more/less valence electrons than the host atoms can lead to increased conductivity through $n$-/$p$-type doping, while magnetic dopants could show increased interactions between carriers and spins due to the confinement within the NC, promising for spintronic devices [8]. So far, efforts to utilize dopants for achieving exciton localization have been limited to protecting materials against photooxidation by suppressing degradation reactions at the NC surface under prolonged illumination, *e.g.* in solar cells [9]. In these systems, efficient energy transfer to the dopant results in complete quenching of the host exciton emission. Alternatively, attempts to exploit exciton localization to increase emission rates rely on core-shell architectures for wave function engineering [10,11], which are synthetically challenging [12].

Using transient optical spectroscopy and *ab initio* calculations, in this letter we report that manganese doping in perovskite NCs results in an increase in radiative exciton recombination rate, which we attribute to an increased degree of exciton localization and concomitant overlap of electron and hole wave functions. Our results present a fundamental mechanism of how transition-metal doping in perovskites impacts band structure and excitonic properties of metal-halide perovskite nanocrystals, and provide a new concept for application in high-performance LEDs, as well as related fields like photocatalysis.

**Results**

We first characterize the synthesized Mn-doped $CsPb(Cl,Br)_3$ perovskite NCs with regard to their structural and optical properties (see Supplemental Material [13] for synthesis according to reported protocols [4]). We employ transmission electron microscopy (TEM) and confirm a cubic structure of the NCs with an average size of approximately 12 nm [Fig. 1(a)].

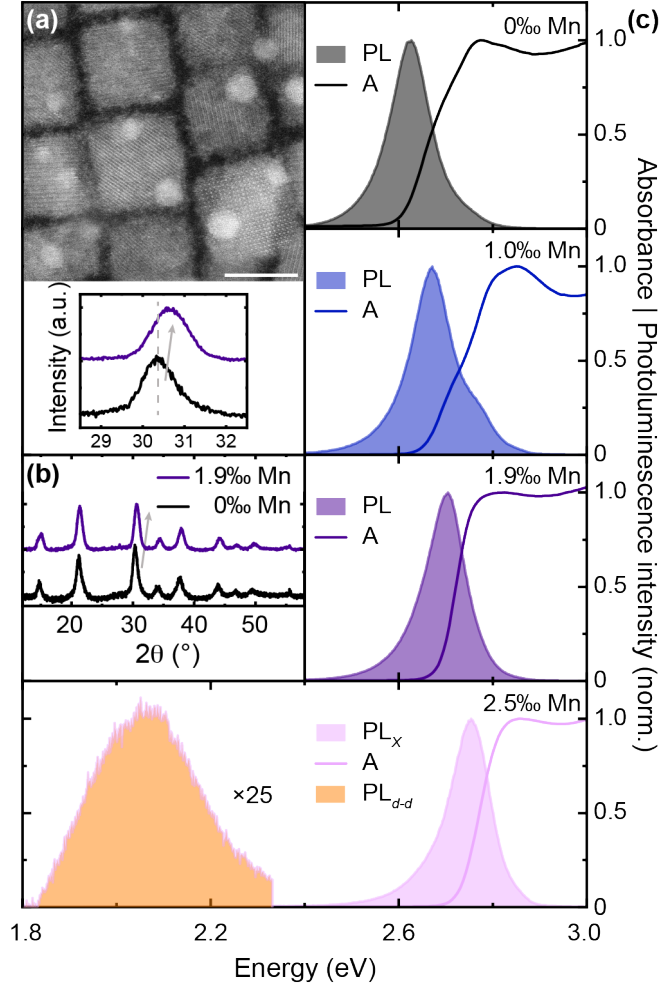

**FIG. 1. Structural and optical properties of manganese-doped perovskite nanocrystals.** (a) Transmission electron microscopy image of doped NCs (shown exemplary for 1.9 atomic ‰ Mn:Pb) with cubic morphology and an average crystal size of 12±2 nm. Scale bar is 10 nm. (b) X-ray diffractogram of undoped (black) and 1.9‰-Mn doped (purple) nanocrystals, respectively, showing a slight lattice contraction upon doping. (c) Steady-state absorbance (bold lines) and photoluminescence (PL, filled) of NC solutions for increasing Mn-doping, showing a doping-induced blue-shift of the excitonic transition around 2.7 eV. Samples were photoexcited with 3.1 eV pulsed excitation at a fluence of 127 μm cm$^{-2}$. The spin-forbidden manganese $d$-$d$ transition around 2.1 eV is absent in the absorption spectra and only emerges in PL at our highest doping level (see Fig. S1 [13] for full-range spectra).

Similar sizes are found for all undoped and manganese-doped crystals studied here. This allows us to compare the impact of Mn-doping on optoelectronic properties quantitatively, excluding effects related to dielectric screening and quantum confinement that might arise from different crystal shapes or sizes [14–16]. We further take X-ray diffraction data [Fig. 1(b)] on undoped NC films and those doped with 1.9‰ Mn:Pb atomic ratio, as determined by inductively-coupled plasma mass-spectrometry (ICPMS). We find very similar diffraction peak patterns across all compositions, with the peaks shifting to higher angles with increasing Mn-doping, as reported before [17–19]. This relates to a moderate degree of lattice contraction which is expected from the incorporation of the smaller manganese(II) ion (1.4 Å) substituting the larger lead(II) ion (1.8 Å) in the octahedral halide coordination sphere. In the steady-state absorption and photoluminescence (PL) spectra [Fig. 1(c)] we observe a gradual blue-shift in both absorption and emission with increasing doping level, ranging from 0 to 2.5‰, consistent with previous reports on manganese-doped perovskite NCs [18–20]. All compositions show an excitonic peak close to the absorption onset. Notably, while all samples exhibit an intense blue excitonic emission ($X$) at around 2.7 eV, only the most doped 2.5‰ sample shows a low-intensity broad orange emission centred at 2.1 eV. This spectral feature is the well-known manganese(II) $^4T_1 \longrightarrow {}^6A_1$ (*d-d*) transition. Due to its spin- and parity-forbidden nature, this transition is not detected in our absorbance spectra, and its weak PL intensity reflects on its low oscillator strength. This is further confirmed by the long emission lifetime on the order of tens of microseconds (see Fig. S2 [13]). Most notably, the excitonic emission increases by a factor of 3.3 from 0 to 1.9‰ manganese doping, and then drops again for the 2.5‰ sample (see also Fig. S1 [13]). This finding is important for optimising colour-pure blue light-emitting diode (LED) efficiencies – a wavelength regime in which it is particularly hard to achieve high efficiencies [21–25]. Our results are in contrast to the emission of Mn-doped perovskite NCs reported previously [18–20,26], in which the excitonic emission remained mostly unchanged or decreased upon Mn doping. In those studies, the manganese doping levels are often on the order of percent [17–19,27,28] and in consequence the manganese emission competes with similar or even higher intensity with the excitonic emission. The doping levels we employ here are an order of magnitude lower than those reported systems, which directly translates into a cheaper manufacturing route to these optimised devices. In the following we will investigate why the efficiency of the excitonic emission increases.

In Figure 2, we quantify the positive impact of the Mn-doping on the emission properties.

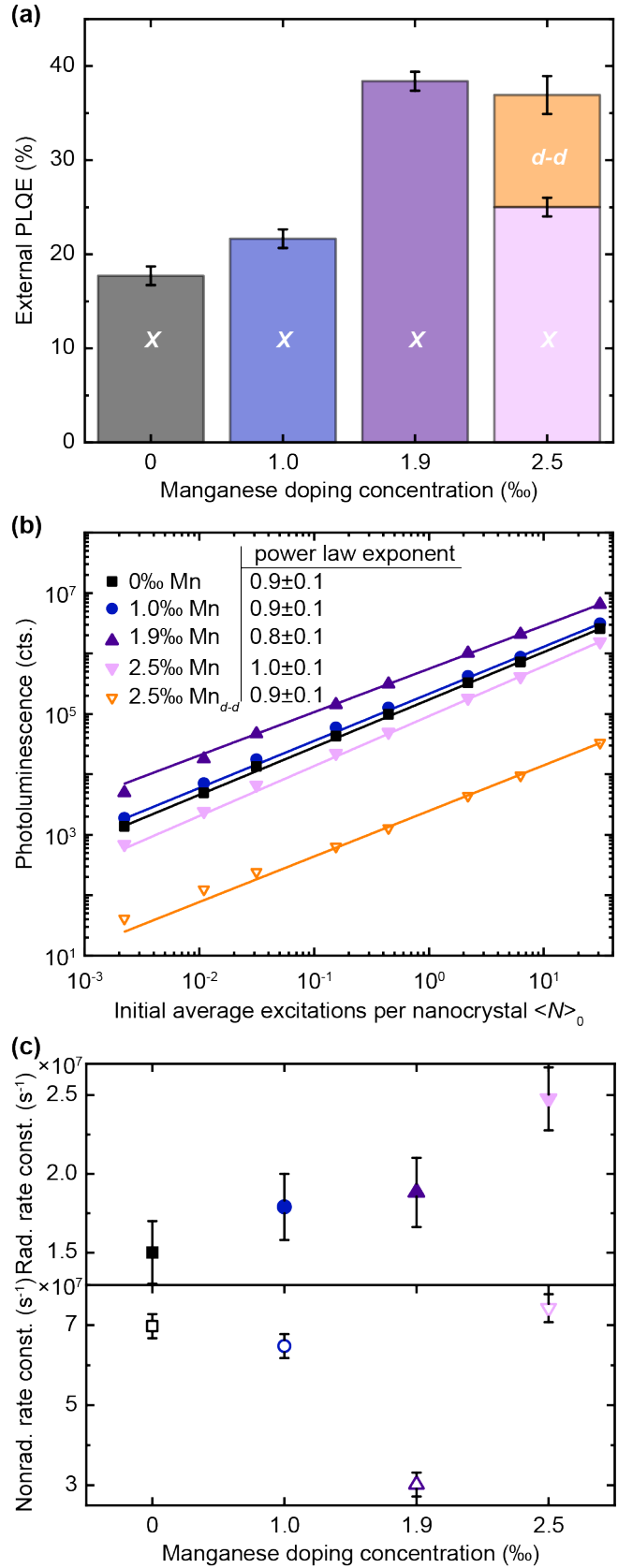

**FIG. 2. Impact of manganese doping on radiative recombination.** (a) Photoluminescence quantum efficiency (PLQE) as function of manganese doping for the perovskite excitonic (*X*) and Mn (*d-d*) emission with a maximum for 1.9‰ Mn-doping. (b) Fluence-dependence of spectrally integrated PL intensity over exciton and Mn emission, respectively. Solid lines are power-law fits of $P^m$ with excitation power $P$ and slope $m$ indicated in the figure panel. All series exhibit linear dependence,

indicating excitonic emission. (c) Dependence of radiative and non-radiative recombination rates on Mn doping, showing a continuous increase in radiative rate and a minimum in non-radiative rate at 1.9‰ Mn doping. Samples were photoexcited with 3.1 eV pulses (repetition rate 1 kHz, pulse duration ~100 fs).

We find that the PLQE of the excitonic emission increases by a factor of 2.2 upon Mn doping up to 1.9‰ manganese, reaching values around 40% [Fig. 2(a)]. For the highest doping level of 2.5‰, the PLQE of the excitonic emission decreases again, while the Mn *d-d* emission becomes detectable. This suggests the existence of a threshold doping level between 1.9 and 2.5‰, upon which energy transfer from the perovskite host exciton to the dopant becomes significant.

Next, we investigate the fluence-dependence of the luminescence [Fig. 2(b)]. We find that the photoluminescence of the excitonic emission of all compositions investigated, as well as the *d-d* emission displayed by the highest doping concentration, scale linear over the studied range of excitation fluences. This indicates that the time-integrated emission is dominated by single-exciton recombination even in the case of high excitation levels for which several excitons are generated initially in a NC [14]. The excitonic behaviour agrees with the pronounced excitonic absorption peak and weak quantum confinement [14] of excitons with predicted Bohr radius of about 2 nm in our NCs with size of about 12 nm.

By employing low-fluence time-resolved single-photon counting experiments (TCSPC, Fig. S2 [13]) we further find that the 1.9‰ composition exhibits the longest average PL lifetime. Combining PLQE and TCPSC results, we can extract the radiative and non-radiative recombination rates of all compositions [Fig. 3(c), see Supplemental Material [13] for calculations]. The non-radiative rate, which represents loss channels for luminescence, decreases upon Mn-doping, reaching its minimum for the composition with the highest PLQE. This is indicative of trap passivation, most likely filling of halide vacancies by the additional halide ions introduced through the manganese salt, that has been claimed as the main reason for the increased performance with Mn doping so far [4,29]. Notably, we further find that the radiative recombination rate increases with doping level, leading to the highest radiative rate of $2.5 \times 10^7$ s$^{-1}$ for the 2.5‰ doping. This increase in the radiative recombination rate, which relates to intrinsic properties of a material, indicates that the manganese dopants alter in fact the exciton dynamics and electronic structure in the perovskite host.

To identify the fundamental origin of the increased radiative recombination rates, we track the time-dependent excited-state dynamics by comparing kinetics obtained from the excitonic PL and the ground-state bleach (GSB) kinetics of transient absorption (TA) under the same excitation conditions [Fig. 3(a), see Fig. S3 [13] for full TA map].

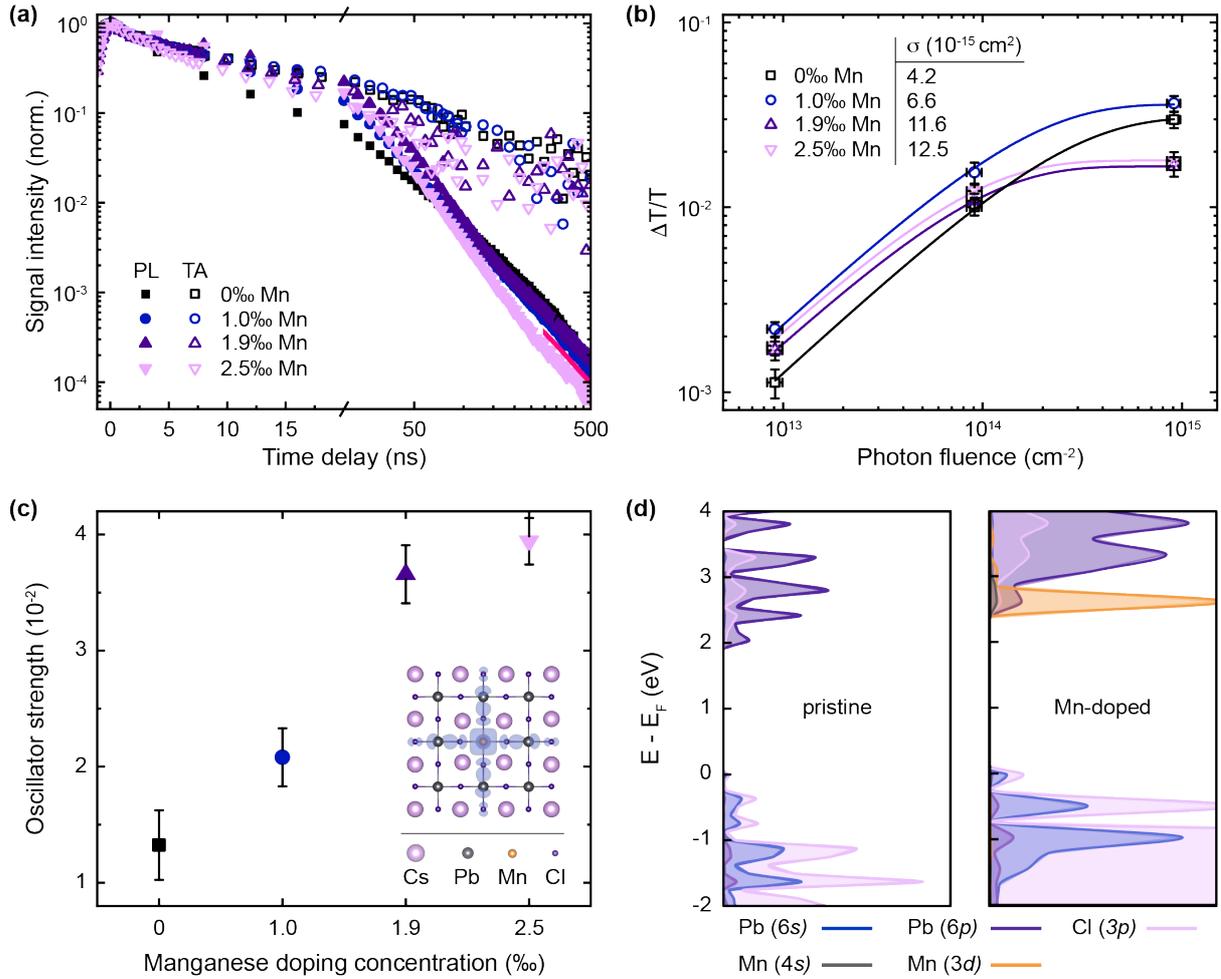

**FIG. 3. Charge carrier dynamics and localization in manganese-doped perovskite nanocrystals.** (a) Transient photoluminescence (PL) and transient absorption (TA) ground-state bleach (GSB) kinetics, showing similar lifetimes in TA for all compositions, while PL initially remains higher for doped samples. Monoexponential kinetics are observed for time delays beyond 50 ns for all compositions with a shared lifetime of 116±2 ns (red line). PL and TA signals were spectrally integrated over the respective peak maximum; initial excitation density approx. 1.1 excitations per NC on average. (b) Initial GSB within the ns-resolution of the experiment as a function of incident photon fluence. The absorption cross-section values $\sigma$ extracted from fits (solid lines, see main text for details) increase with doping level. All samples excited at 3.1 eV for (a) and (b). (c) Values of the oscillator strength per unit cell of the band-to-band transitions determined from experimental absorption cross-sections (see main text for details). Inset: Frist-principles calculation of the charge density for Mn-doped perovskite, showing carrier localization at manganese dopant. Value at isosurface is $2.36 \times 10^{-3}$ $e$ Å$^{-3}$. (d) Projected density of states for CsPbCl$_3$ (left) and CsPb$_{0.963}$Mn$_{0.037}$Cl$_3$ (right), both calculated from first principles with the PBEsol functional including 20% Hartree-Fock exchange. We find significant hybridization of the Mn 4$s$ states with the host which leads to charge localization, responsible for the observed oscillator strength increases.

Both techniques (TA and PL) probe the decay of photo-generated excitons. TA kinetics probe how many excitations have not yet decayed to the ground state, i.e. all remaining excitations, including those which cannot contribute to PL anymore. The PL signal represents only the amount of radiatively recombined carriers per unit time. However, the PL *kinetics* also reflect non-radiative losses [16,17,30]. This explains why all PL kinetics generally decay faster than the respective TA kinetics. In more detail, within the initial few ns the PL matches closely the TA signal. In this regime we expect mostly radiative exciton recombination to take place because (i) the ns-lifetimes characteristic for this process are observed and (ii) charge-neutral exciton traps, which would be the only other scenario where the PL kinetics would fully track the TA decay, are unlikely to play a major role in these high-PLQE materials. This predominantly radiative regime increases from about 5 ns for the undoped to about 15 ns for the highest doping concentration. For the longer time delays beyond 50 ns all compositions follow a slow monoexponential decay with similar lifetimes (see Fig. S3 [13] for individual fits). We assign this long-lived component to non-radiative trap-assisted recombination [31,32]. The longer PL lifetimes upon doping indicate a reduction in non-radiative decay. The TA kinetics decay faster for the doped samples up to 50 ns, which is unexpected for reduced non-radiative recombination rates and a strong further indication for an increase in radiative rates. For time delays beyond 50 ns all samples show similar TA decays with longer lifetimes than the PL decays, suggesting a trap-dominated regime with long relaxation times and low radiative efficiency.

In order to quantify the underlying mechanism of enhanced radiative rates we next determine the oscillator strengths of the transitions from the absorption cross-sections. We first calculate the average excitations per NC $\langle N \rangle$ from two independent methods yielding similar values (see Supplemental Material [13] for details on both methods): one is a bulk-estimate based on the NC concentration and steady-state absorbance, while the other one is based on the assumption of a Poisson distribution of initial excitations in the NC ensemble. The average excitations per NC are directly related to the absorption cross-section $\sigma$ via $\langle N \rangle = \sigma j$. For this, following the approach demonstrated by Klimov [16,33] and others [34,35], we take the probability $p$ of a NC to contain $i$ excitations to be $p_i = \langle N \rangle^{i/i!} e^{-\langle N \rangle}$. After multi-exciton recombination is completed at very early times, which we confirm by extracting biexciton lifetimes of 10-25 ps (see Fig. S5 [13] for ultrafast TA data, as well as the literature [17]), each NC cannot contribute to more than a single de-excitation event to be detected. Hence, when plotting the GSB intensity (after initial cooling and biexciton decay) as a function of photon fluence $j$ (number of photons incident per cm$^2$), we observe a saturation behavior at high fluences [Fig. 3(b)]. This fluence-dependence can be described by a Poisson distribution with the GSB intensity being proportional to $1 - p_0 = 1 - e^{-\langle N \rangle}$ [33,36]. From our experimental TA data we find the absorption cross-section to increase with Mn-doping level, yielding a nearly three-fold increase in the $\sigma$ value for the highest doping concentration. With the measured absorption cross-section, we can then directly access the oscillator strength $f$ for the band-to-band transition [Fig. 3(c)] employing a modified version of the Strickler-Berg relation [37]:

$$f = \frac{8m_0 0.2303\, n_m c^2 \varepsilon_0 \pi}{e^2 \lambda |F|^2} \sigma \qquad (1)$$

where $m_0$ is the free electron rest mass, $n_m$ the refractive index of the medium, here toluene, $c$ is the speed of light in a vacuum, $\varepsilon_0$ is the vacuum permittivity, $e$ is the elementary charge and $\lambda$ is the wavelength of the optical transition. $F = 3\varepsilon_m/(\varepsilon_s + 2\varepsilon_m)$ is the local field factor to account for the screening of the nanoparticle modelled as a sphere, with $\varepsilon_m$ and $\varepsilon_s$ being the dielectric constants of the medium and semiconductor, respectively. Since the oscillator strength quantity originates from a single-oscillator model, the values obtained were divided by the ratio of unit cell volume to NC volume before plotting, though the sum following the Thomas-Reiche-Kuhn rule could also be used [38,39]. We observe a threefold increase in the oscillator strength for the most doped sample. We find very similar values for the oscillator strength from the measured radiative rates, thus confirming our finding from two different sets of experiments, fundamentally connected via the Einstein relations [40]. The determined increase in oscillator strength, which is a direct measure of the dipole matrix element of the transition, can only be explained by two changes to the system upon doping: (i) changes in the dielectric constant of the material, which is highly unlikely given that only 60-80 atoms are substituted in our permille doping regime per nanocrystal (~$5 \times 10^5$ atoms); or (ii) the electron-hole overlap $\Theta_{e-h}$ increases, which is given by [41]:

$$\Theta_{e-h} = |\int \psi_e^*(r)\psi_h(r)dV|^2 \quad (2)$$

where $\psi_{e,h}(r)$ are the electron and hole envelop wavefunctions, respectively, which are integrated over the volume $V$. The local distortion of the perovskite lattice induced by the Mn dopants is likely to break the lattice periodicity in the interior of the NC and create localization sites for the photo-generated excitons, which leads to higher values for the electron hole overlap and thus probability of radiative decay. Assuming no changes to the dielectric constant, we find $\Theta_{e-h}$ to increase by 158% for the highest doping level (see Supplemental Material [13] for details and refs. [41,42]).

To further probe the proposed influence of Mn-doping on exciton recombination, we perform electronic structure calculations using density functional theory (DFT, see Supplemental Material [13] for computational details). Via the supercell approach, we model bulk pristine and Mn-doped $CsPbX_3$ compositions for X = Cl, Br or a mixture thereof. We first confirm a direct band gap at the R-point of the Brillouin zone, which upon doping increases by ~0.2 eV in agreement with the experimentally determined values. In line with previous reports [43,44], the valence band maximum (VBM) of the undoped composition is mainly composed of antibonding hybrids from Pb 6*s* and halide *p* orbitals, while the conduction band minimum (CBM) consists mainly of empty Pb 6*p* orbitals. However, once the Pb is partially replaced with Mn in the doped compositions, the VBM is perturbed, resulting in a band that mostly resembles the energetically lower lying isolated halide *p* orbitals. At the same time, the perturbation in the periodicity of the Pb 6*p* orbitals leads to the destabilization of the CBM compared to the undoped case. Both perturbations hence contribute towards the observed widening of the bandgap as a consequence of mostly electronic rather than structural changes. Importantly, these perturbations also reduce the dispersion of both VBM and CBM, indicating a more localized electron and hole state with higher effective masses in the doped case.

To identify the element-specific changes that occur through the substitution of Pb for Mn, we calculate the real-space charge distribution of the CBM for pristine and Mn-doped $CsPbCl_3$ using the PBEsol functional with 20% additional Hartree-Fock exchange [Fig. 3(c), inset]. We observe that the charge distribution shows a more localized character compared to the undoped case (see Figs. S8 and S9 [13]) with the largest coefficients in proximity to the central Mn atom and along the Mn-Cl-Pb bonds, while the contributions from the remaining perovskite scaffold become negligible. This charge localization corresponds to an increased electron hole wavefunction overlap and thus increases the probability of radiative recombination [45] upon doping, observed experimentally. The origin of this localization effect is found in the projected density of states [pDOS, Fig. 3(d)]. Here, the Mn $4s$ states as well as the Mn $3d$ states energetically coincide with the CBM of the perovskite but intriguingly, while the $3d$ orbitals have a negligible effect on the band edge, the $4s$ orbitals strongly hybridize with it and thus modify the host wavefunction significantly. This significant hybridization of the Mn $4s$ states with the host leads to charge localization, responsible for the observed radiative rate increases, while the Mn $3d$ states do not mix with the CBM and hence form a competing decay channel at higher concentrations.

We note that the hybridization of the Mn $4s$ orbitals with the host shown here for the Mn-doped pure-chloride composition is halide-dependent (see Fig. S7 [13]): With increasing Br concentration the unoccupied perovskite states shift towards lower energies, while the Mn $4s$ states remain invariant. This off-set of the relative energies causes a reduced hybridization. The resulting loss in localization could explain why similarly beneficial doping effects on the radiative rate of the perovskite NCs have not been reported in the literature for pure-bromide compositions so far.

Further, we observe that the relative position of the vacant Mn $3d$ states is not only dependent on the Mn concentration but on the halide content, too (see Fig. S7 [13]): With increasing Br content, the unoccupied Mn $3d$ orbitals are pushed towards higher energies, such that $d$-$d$ transitions that were previously lying within the host band gap become inaccessible as an alternative decay pathway. This suggests that a balanced halide ratio and distribution is beneficial in order to suppress this host exciton decay channel. We conclude that both the Mn *and* the halide concentration are levers that can be used to tune localization and thus PLQE. This further points towards MnCl$_2$ acting not only as a means to fill halide vacancies and reduce trap densities that cause non-radiative decay, but more importantly to engineer the exciton wave function for optimized radiative yields through the halide ratio and distribution of Mn centers.

Lastly, we note that the use of other transition metals as dopants could be a promising avenue to tackle the challenges mentioned above: By tuning the element-dependent degree of *s*-orbital hybridization and mitigating the formation of *d*-states *via* the choice of closed-shell ions, the localization effect we observe can be exploited to maximize radiative rates.

**Conclusions**

We report that manganese-doping of perovskite nanocrystals increases their luminescence yields due to an increased radiative recombination rate of excitons and further reduction of non-radiative losses through defects. We can attribute the enhanced luminescence rates to increased oscillator strength from stronger exciton localization, with a concomitant increase in overlap of electron and hole wavefunctions. Our results demonstrate that transition-metal doping provides detailed control of electronic structure and exciton dynamics in metal-halide perovskite nanocrystals and opens a route towards very efficient light-emitting devices.

**Acknowledgements**

S.F. acknowledges funding from the Studienstiftung des deutschen Volkes and EPSRC, as well as support from the Winton Programme for the Physics of Sustainability. T.W. acknowledges funding from EPSRC NI grant agreement EP/R044481/1. D.N.C. and M.K.G. acknowledge the support of the Rowland Fellowship at the Rowland Institute at Harvard University. I.B., B.P. and B.M. acknowledge support from the Winton Programme for the Physics of Sustainability. B.M. also acknowledges support from the Gianna Angelopoulos Programme for Science, Technology, and Innovation. Part of the calculations was performed using resources provided by the Cambridge Tier-2 system operated by the University of Cambridge Research Computing Service funded by EPSRC Tier-2 Capital Grant No. EP/P020259/1. I.B. thanks Yun Liu for providing the mixed-halide crystal structure. R.H.F. acknowledges support from the Simons Foundation (grant 601946). F.D. acknowledges funding from the Winton Programme for the Physics of Sustainability and the DFG Emmy Noether Program.

**Author contributions**

S.F. and F.D. conceived and planned the experiments with additional input from T.W., R.H.F. and D.C. M.G. and T.N. fabricated samples. M.G. performed TEM and XRD measurements. S.F. performed the optical measurements and analysed all data. I.B. and B.P. performed the DFT calculations with input from B.M. S.F. wrote the manuscript and compiled figures, with discussion of results and feedback on the manuscript from all authors.

**Competing interests**

The authors declare no competing interests.